\documentclass[journal]{IEEEtran}
\IEEEoverridecommandlockouts

\usepackage{microtype}
\usepackage{cite}

\ifCLASSINFOpdf
\usepackage[pdftex]{graphicx}
\DeclareGraphicsExtensions{.pdf,.jpeg,.png,.svg}
\else

\fi

\ifCLASSOPTIONcompsoc
\usepackage[caption=false,font=normalsize,labelfont=sf,textfont=sf]{subfig}
\else
\usepackage[caption=false,font=footnotesize]{subfig}
\fi

\usepackage{amsmath}
\usepackage{amssymb}
\usepackage{amsfonts}
\usepackage{bm}
\usepackage{mathtools}
\usepackage{pifont}
\usepackage[dvipsnames]{xcolor}

\usepackage{tikz}
\usepackage{tikzscale}
\usepackage{pgfplots}
\usetikzlibrary{shapes,fit,arrows,calc,positioning,angles,quotes}

\usetikzlibrary{calc,patterns,angles,quotes}
\usetikzlibrary{decorations}
\usetikzlibrary{decorations.pathmorphing,decorations.markings,decorations.pathreplacing}
\usetikzlibrary{positioning, shapes, arrows, calc, backgrounds, angles, quotes, intersections,fit}
\usetikzlibrary{arrows.meta}
\usetikzlibrary{external}

\tikzstyle{every node}=[font=\footnotesize]
\pgfplotsset{compat=1.15}
\pgfplotsset{label style={font=\footnotesize},
tick label style={font=\footnotesize},
legend style={font=\footnotesize},
xlabel shift = -2 pt}

\tikzset{>={Latex[scale=0.65]}, transform shape}
\tikzstyle{very densely dashed} = [dash pattern=on 1.5pt off 1pt]

\DeclareMathOperator*{\argmin}{arg\,min}

\newcommand\centerofmass{%
	\tikz[radius=0.15cm] {%
		\fill (0,0) -- ++(0.15cm,0) arc [start angle=0,end angle=90] -- ++(0,-0.3cm) arc [start angle=270, end angle=180];%
		\draw[semithick] (0,0) circle;%
	}%
}

\usepackage{booktabs}
\usepackage{epsfig}
\usepackage{float}

\usepackage[hidelinks]{hyperref}
\usepackage{cleveref}
\Crefname{figure}{Fig.}{Figs}
\crefname{figure}{Fig.}{Figs}
\Crefname{Table}{Table}{Tables}
\crefname{table}{Table}{Tables}

\begin{document}

		\title{Uncertainties in Robust Planning and Control of \\ Autonomous Tractor-Trailer Vehicles}
		
		\author{Theodor Westny\IEEEauthorrefmark{1}, Bj\"orn Olofsson\IEEEauthorrefmark{2}\IEEEauthorrefmark{1}, and Erik Frisk\IEEEauthorrefmark{1} 					
		\thanks{\scriptsize\IEEEauthorrefmark{1}Department of Electrical Engineering,
			Linköping University, Sweden.}
		\thanks{\scriptsize\IEEEauthorrefmark{2}Department of Automatic Control,
			Lund University, Sweden.}
		}
		\maketitle

		\begin{abstract}
To study the effects of uncertainty in autonomous motion planning and control, an 8-DOF model of a tractor-semitrailer is implemented and analyzed.
The implications of uncertainties in the model are then quantified and presented using sensitivity analysis and closed-loop simulations.
The analysis reveals that the significance of various model parameters varies depending on the specific scenario under investigation. 
By using sampling-based closed-loop predictions, uncertainty bounds on state variable trajectories are determined.
Our findings suggest the potential for the inclusion of our method within a robust predictive controller or as a driver-assistance system for rollover or lane departure warnings.
\end{abstract}

		\IEEEpeerreviewmaketitle

		\section{Introduction}
Important remaining challenges for autonomous vehicle research are associated with unknowns originating from various sources of uncertainties---be it from the surrounding environment~\cite{fors2022resilient, westny2023graph, zhou2023interaction}, or from the ego-vehicle itself.
This research primarily focuses on the latter aspect, which encompasses potentially unknown internal parameters or unmodeled dynamics of the ego-vehicle.
Investigating tractor-trailer vehicles for these purposes is interesting for several reasons.
Compared to conventional passenger cars, tractor-trailers have complex dynamics due to factors like increased mass, a higher center of gravity, and unstable yaw modes. These characteristics limit maneuverability and the range of viable control inputs~\cite{sampson2000active, tai1999robust}.
Moreover, commercial trucks frequently operate with different trailers, meaning, they may not be equipped with sensors to measure the trailer's states.
Even if that is not the case, some of the most significant model parameters, e.g., the moment of inertia, may be difficult to measure or estimate~\cite{sadeghi2019robust}.
Given these complexities, there's a clear need for predictive models and methods that effectively manage uncertainties, especially in the context of autonomous driving applications. This research aims to address this gap, proposing solutions that can enhance the reliability and safety of self-driving tractor-trailer vehicles.
		\subsection{Background}
\label{sec:prob-desc}
The choice of vehicle model is determined by its intended use and is influenced by implementation and computational limitations, often leading to reduced-order dynamic or kinematic models \cite{kong2015kinematic, islam2019well}.
Such models, while efficient, inherently introduce uncertainties in model-based control systems. These uncertainties can arise from a variety of factors, including but not limited to unknown tire properties and varying environmental conditions.
Understanding and quantifying the impact of these uncertainties is crucial for ensuring the safety and effectiveness of autonomous driving systems across diverse scenarios. This research is dedicated to this very purpose.
While there exist numerous potential sources of uncertainties in a predictive model, our study narrows its focus to a select set of parameters specifically related to the trailer.
These parameters were identified through a comprehensive sensitivity analysis of the system, allowing us to concentrate our efforts on the most impactful factors.

\subsection{Contributions}
In this paper, methods for uncertainty quantification of an autonomous tractor-semitrailer vehicle are developed.
We adopt modeling principles from the literature, contrasting many prior works on robust control by opting for high-complexity nonlinear dynamic models. 
The main design choices concern the specific vehicle configuration and the modeling of external forces.
The model is used to conduct a sensitivity analysis of the system with respect to several model parameters belonging to the trailer.
By defining a distribution over model parameters and using sampling-based strategies, we show how to compute state variable trajectory boundaries under parameter uncertainties.
Finally, we show how our method can be used as an early-warning system for potential rollover and lane departure.

		\section{Model}
\label{sec:model}
\begin{figure}[t]
	\centering
	\resizebox{0.95\columnwidth}{!}{%
		\begin{tikzpicture}[transform shape]
  
  \tikzset{wheel/.style = {draw, circle, minimum size=6mm, thick}}
    
  \tikzstyle{coil}=[semithick,decorate,decoration={coil,aspect=0.5,pre length=1mm,post length=0.1cm,segment length=1.8,amplitude=1.5,pre=curveto,post=curveto}]
  
  \tikzstyle{classic_spring}=[thick,decorate,decoration={zigzag,pre length=0.05cm,post
 length=0.05cm,segment length=5}]
  
  \tikzstyle{damper}=[semithick,decoration={markings,  
   mark connection node=dmp,
   mark=at position 0.35 with 
   {
     \node (dmp) [semithick,inner sep=0pt,transform shape,rotate=-90,minimum
 width=3pt,minimum height=1pt,draw=none] {};
     \draw [semithick] ($(dmp.north east)+(1pt,0)$) -- (dmp.south east) -- (dmp.south
 west) -- ($(dmp.north west)+(1pt,0)$);
     \draw [semithick] ($(dmp.north)+(0,-1pt)$) -- ($(dmp.north)+(0,1pt)$);
   }
 }, decorate]
 
 \newcommand{\AxisRotator}[1][rotate=0]{%
    \tikz [x=0.25cm,y=0.60cm,line width=.15ex,-stealth,#1] \draw (0,0) arc (-150:150:0.4 and 0.44);%
}

  \coordinate (z_hat) at (0, 1.25);
  \coordinate (y_hat) at (0.75, 0.75);
  \coordinate (x_hat) at (1.25, 0);
    
  \coordinate (cog_h) at (0, 0.5);
  \coordinate (rz) at (0, 0.7);
  \coordinate (rz_h) at (0, 0.35);
  
  \coordinate (Ls) at (-8.5, 0);
  
  \coordinate (Lt2) at (-4.8, 0);
  \coordinate (Lt1) at (-2, 0);
  
  \coordinate (Tw) at (1, 1);

  \begin{scope}[]
      \node [wheel] at (Ls) (rw6) {};
      \node [wheel] at ($ (rw6.center) + (Tw) $) (rw5) {};
      \draw [very thick] (rw6.center) -- (rw5.center);
      
      \coordinate (r6) at ($ (rw6.center) + (rz) $);
      \coordinate (r5) at ($ (rw5.center) + (rz) $);
      
      \draw[very thick] (r6) -- (r5) node [midway] (semi) {};
      
      \node [wheel] at (Lt2) (rw4) {};
      \node [wheel] at ($ (rw4.center) + (Tw) $)  (rw3) {};
      \draw [very thick] (rw4.center) -- (rw3.center);
      
      \coordinate (r4) at ($ (rw4.center) + (rz) $);
      \coordinate (r3) at ($ (rw3.center) + (rz) $);
      
      \draw[very thick] (r4) -- (r3) node [midway] (rear) {};
      
      \draw[very thick] (r4) -- (r3) node [midway] (rear) {};
      
      \coordinate (below_semi_cog) at ($(semi.center)!0.5!(rear.center)$) {};
      
      \draw[thick] (semi.center) -- (below_semi_cog.center);
      \draw[thick] (below_semi_cog.center) -- ($ (below_semi_cog.center) + (cog_h) $) node [transform shape] (semi_cog) {\centerofmass};
    
      \coordinate (upwards) at ($(below_semi_cog.center)!0.6!(rear.center)$) {};
      
      \draw[thick] (below_semi_cog.center) -- ($ (upwards.center) + (rz_h) $) node[] (downwards) {};
      
      \node [wheel] at (Lt1) (rw2) {};
      \node [wheel] at ($ (rw2.center) + (Tw) $) (rw1) {};
      \draw [very thick] (rw2.center) -- (rw1.center);
      
      \coordinate (r2) at ($ (rw2.center) + (rz) $);
      \coordinate (r1) at ($ (rw1.center) + (rz) $);
      
      \draw[very thick] (r2) -- (r1) node [midway] (front) {};
      
      \coordinate (hitch) at ($(rear.center)!0.35!(front.center)$) {};
      \draw[thick] (hitch.center) -- ($ (hitch.center) + (rz_h) $) node [] (trailer_connection) {};
      \draw[thick] (downwards.center) -- (trailer_connection.center);
      
      \filldraw (trailer_connection.center) circle (0.8pt);

      \coordinate (hitch_roll_start) at ($(rear.center)!0.45!(front.center)$) {};
      
      \coordinate (hitch_roll_end) at ($(rear.center)!0.7!(front.center)$) {};
      
      \draw[classic_spring]  (hitch_roll_start) -> (hitch_roll_end);
      
      \coordinate (below_cog) at ($(rear.center)!0.8!(front.center)$) {};
      
      \draw[thick] (below_cog) -- ($ (below_cog) + (cog_h) $) node [transform shape] (tractor_cog) {\centerofmass};
      
      \draw[thick] (hitch_roll_start) -- (rear.center);
      \draw[thick] (hitch_roll_end) -- (front.center);
      
      \filldraw (hitch.center) circle (2pt) node[] (origo) {};
      \draw [->, semithick, very densely dashed] ($ (origo.center) - (z_hat) $) -- ($ (origo.center) + (z_hat) $) node[left] {\normalsize$z$};

      \draw [->, semithick, very densely dashed] ($ (origo.center) - (x_hat) $) -- ($ (origo.center) + (x_hat) $) node[below left] {\normalsize$x$};
      
      \draw [->, semithick,  very densely dashed] ($ (origo.center) - (y_hat) $) -- ($ (origo.center) + (y_hat) $) node[left] {\normalsize$y$};

      \draw[coil]  (r6) -> (rw6.center);
      \draw[damper] ($(rw6.center)!0.16!(rw5.center)$) ->($(r6)!0.16!(r5)$);
      
      \draw[coil]  (r5) -> (rw5.center);
      \draw[damper] ($(rw6.center)! 1- 0.16!(rw5.center)$) ->($(r6)! 1- 0.16!(r5)$);

      \draw[coil]  (r4) -> (rw4.center);
      \draw[damper] ($(rw4.center)!0.16!(rw3.center)$) ->($(r4)!0.16!(r3)$);
      
      \draw[coil]  (r3) -> (rw3.center);
      \draw[damper] ($(rw4.center)!1-0.16!(rw3.center)$) ->($(r4)!1-0.16!(r3)$);

      \draw[coil]  (r2) -> (rw2.center);
      \draw[damper] ($(rw2.center)!0.16!(rw1.center)$) ->($(r2)!0.16!(r1)$);
      
      \draw[coil]  (r1) -> (rw1.center);
      \draw[damper] ($(rw2.center)!1-0.16!(rw1.center)$) ->($(r2)!1-0.16!(r1)$);
  \end{scope}

\end{tikzpicture}
	}%
	\caption{Schematics of the vehicle configuration used in the paper.}
	\label{fig:vehicle_configuration}
\end{figure}
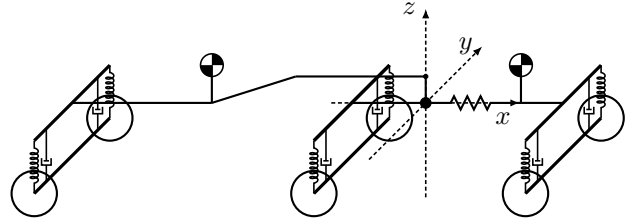

The need of a relevant model of adequate complexity to perform the analysis is essential.
We adopt the methodology in \cite{gafvert20049}, and implement an 8-DOF model of a tractor-semitrailer combination, schematically illustrated in \cref{fig:vehicle_configuration}.
The vehicle configuration was chosen based on its prevalence in commercial applications and its frequent use in works on control \cite{sampson2000active, tai1999robust, gafvert20049, chen2000lateral}.
The model parameters were determined using a parameter-estimation method and simulation data from TruckMaker.
The modeling principle is briefly described in the following section, alongside modifications to the original model.

\subsection{Modeling}
\label{sec:modeling}
The system consists of five moving coordinate systems, two unsprung frames that rotate about the $z$-axis of the inertial frame, and three (front and rear for the tractor) sprung frames that can roll and pitch about the $x$ and $y$-axes of their unsprung respective frame.
The tractor and trailer are connected through the sprung bodies, expressed as a kinematic constraint.
All reference systems coincide and the origin is placed at the hitch.
The choice of origin simplifies the dynamic equations, as it has the unique property of being common for both of the vehicle units (see \cite{gafvert20049} for detailed model equations).
The states in the model are presented in \cref{tab:state_vars}. %
Inputs are the tractor front wheel steering angle $\delta_f$ and the rear tire slip ratios $\kappa$.

\begin{table}[t]
	\caption{State variables.}
	\label{tab:state_vars}
	\centering
	\begin{tabular}{c l}
		\toprule
		State & Description\\
		\midrule
		$v_x$ & Longitudinal velocity of the hitch\\
		$v_y$ & Lateral velocity of the hitch\\
		$x_i$ & COG longitudinal position \\
		$y_i$ & COG lateral position \\ 
		$\dot{\phi}_i$ & Roll rate \\
		$\dot{\theta}_i$ & Pitch rate\\
		$\dot{\psi}_i$ & Yaw rate\\
		$\phi_i$ & Roll angle\\
		$\theta_i$ & Pitch angle\\
		$\psi_i$ & Yaw angle\\
		\midrule
		$i \in \{t, s\}$ & $t$ = tractor, $s$ = semitrailer\\
		\bottomrule
	\end{tabular}
\end{table}

\subsubsection{Aerodynamic Resistance}
The aerodynamic resistance $R_a$ is defined as in \cite{wong2008theory}
\begin{equation}
	R_a = \frac{\rho}{2}c_DA_fv_r^2,
\end{equation}
where $\rho$ is the mass density of the air, $c_D$ is the coefficient of aerodynamic resistance, $A_f$ is characteristic area of the vehicle, and $v_r$ is the speed of the vehicle relative to the wind.
The model includes aerodynamic resistance acting on both of the sprung bodies in the longitudinal direction.

\subsubsection{Rolling Resistance}
The rolling resistance $R_r$ may be expressed by \cite{wong2008theory}
\begin{equation}
	\label{eq:rolling_resistance}
	R_r = f_r F_z,
\end{equation}
where $F_z$ is the normal force at the contact point and $f_r$ is the coefficient of rolling resistance, where
\begin{equation}
	f_r = c_1 + c_2 v_x^2,
\end{equation}
with $c_1$ and $c_2$ constants.
The rolling resistance acts directly on the tires in the unsprung coordinate systems.

\subsubsection{Ground--Tire Interaction}
For each wheel $i$, the lateral slip angle $\alpha_i$ is defined as in \cite{pacejka2012tire},
\begin{align}
    \frac{\sigma}{v_{x,i}}\dot{\alpha}_i + \alpha_i &= -\arctan\left(\frac{v_{y,i}}{v_{x,i}}\right),
\end{align}
where $\sigma$ is the relaxation length, and $\bm{v}_i$ the velocity of the wheel in its local coordinate system. 
While it is common to model the tire forces as a nonlinear dependence on the slip, the ground--tire interaction forces for heavy vehicle tires tend to be more linear than their light-vehicle counterparts \cite{pacejka2012tire}. 
This property is utilized, yielding the following ground--tire interaction models:
\begin{subequations}
\begin{align}
    F_{x,i} &= C_\kappa \kappa_i, \\
    F_{y,i} &= C_\alpha \alpha_i,
\end{align}
\end{subequations}
where $C_\kappa$ and $C_\alpha$ are the longitudinal and lateral (cornerning) tire stiffnesses, respectively.

\subsection{Model Parameterization}
\label{sec:model-parameterization}
The medium-complexity model described in \Cref{sec:modeling} is implemented with accessible model parameters and equations, while at the same time having good predictive abilities in selected aspects.
To match the behavior to those of the high-fidelity models in TruckMaker, the parameters in the 8-DOF model were estimated using a \emph{prediction error method} (PEM) \cite{ljung1999system}. Consider a state-space description of the vehicle model,
\begin{subequations}
	\begin{align}
		M(\bm{x}, \bm{p}) \dot{\bm{x}} &= f(\bm{x}, \bm{u}, \bm{p}, t), \\
		\bm{y} &= h(\bm{x}, \bm{u}, \bm{p}, t) + \bm{e},
	\end{align}
\end{subequations}
where $M(\cdot)$ is an invertible matrix, $\bm{x}$ denotes the state vector, $\bm{u}$ the input, $\bm{p}$ the parameter vector, $\bm{y}$ the system output, and $\bm{e}$ measurement noise.
A predicted output of the system, here denoted $\hat{\bm{y}}$, can then be formulated as
\begin{subequations}
	\begin{align}
		M(\bm{x}, \bm{p}) \dot{\hat{\bm{x}}} &= f(\hat{\bm{x}}, \bm{u}, \bm{p}, t), \\
		\hat{\bm{y}} &= h(\hat{\bm{x}}, \bm{u}, \bm{p}, t),
	\end{align}
\end{subequations}
where $\hat{\bm{x}}$ represents the estimated state vector. 
The parameter estimates are then obtained by minimizing a loss function involving the predicted outputs and the measured outputs:
\begin{align}
	\hat{\bm{p}} &=\argmin_{\bm{p}} V_N(\bm{p}), \\
	V_N(\bm{p}) &= \sum_{t=1}^N ||\bm{y}(t, \bm{p}) - \hat{\bm{y}}(t, \bm{p})||^2.
\end{align}
Using the PEM, the cornering stiffnesses, friction coefficients, and coefficients governing the suspension dynamics were determined, yielding satisfactory results with over $95$\% prediction accuracy across several scenarios.

		\section{Sensitivity Analysis}
\label{sec:sens-analysis}
Sensitivity analysis aims to formulate the derivative of the investigated system model with respect to each parameter \cite{maly1996numerical}.
The sensitivity approximations can be used as indicators in determining the importance of nominal parameter value accuracy.
The system of ODEs describing the 8-DOF model is formulated in the form
\begin{equation}
	M(\bm{x}, \bm{p}) \bm{\dot{x}} = f(\bm{x}, \bm{u}, \bm{p}, t).
\end{equation}
Implementing the model using a symbolic framework is favored due to its modularity.
However, to retrieve the states $\bm{x}$, the matrix $M(\cdot)$ needs to be inverted, which symbolically proved infeasible. 
These issues were circumvented by reformulating the model into the implicit form
\begin{equation}
	F(\bm{x}, \dot{\bm{x}}, \bm{u}, \bm{p}, t) \coloneqq M(\bm{x}, \bm{p}) \bm{\dot{x}} - f(\bm{x}, \bm{u}, \bm{p}, t) = 0.
\end{equation}
As such, the equations governing the sensitivity analysis procedure may be formulated as in \cite{maly1996numerical}:
\begin{subequations}
	\begin{align}
		F(\bm{x}, \dot{\bm{x}}, \bm{u}, \bm{p}, t) &= 0, \\
		\frac{\partial F}{\partial \bm{x}} \bm{s}_i	+ \frac{\partial F}{\partial \dot{\bm{x}}} \dot{\bm{s}}_i + \frac{\partial F}{\partial \bm{p}_i} &= 0, \quad i=1,\dots, n_p,
	\end{align}
	\label{eq:sens-analysis}
\end{subequations} 
where $\bm{s}_i = \text{d} \bm{x}/\text{d} \bm{p}_i$ are the sensitivities and $n_p$ is the number of parameters.
In contrast to $M(\cdot)$, the Jacobians $\partial F/\partial \bm{a}$ may be efficiently computed using a symbolic framework.
We used numerical differentiation formulas implemented in \texttt{ode15s} in \textsc{Matlab} \cite{shampine1997matlab} to solve (\ref{eq:sens-analysis}).
		\section{Closed-loop Prediction}
\label{sec:cl-pred}
To investigate how parameter perturbations affect state predictions in closed-loop, we adopted simple but proven methods for autonomous control.
\subsection{Steering Control}
For steering control, we implemented a pure pursuit algorithm. %
The method has a long history of success in the robotics community, much attributed to its simplicity and satisfactory performance \cite{paden2016survey}.
The algorithm works by calculating the curvature that would steer the vehicle to a goal position on the reference path.
The goal position is determined using the vehicle's current configuration and a lookahead distance. %
We use a variable lookahead that depends on the vehicle's target speed:
If the vehicle speed is large, the lookahead should be large such that the goal is far away and the steering-angle magnitude limited.

\subsection{Speed Control}
We used a feedback slip controller in the form
\begin{equation}
	\kappa(t) = K_p e(t) + K_i \int_0^t e(\tau)d\tau,
\end{equation}
where $e(t)$ is the error in speed at time $t$.
The constants $K_p$ and $K_i$ were determined through simulations.
The speed controller is outfitted with anti-windup functionality that limits the size of the integral term.
		\section{Results \& Discussion}
A series of investigations were conducted for both the sensitivity analysis and the closed-loop prediction.

\subsection{Sensitivity Analysis}
The sensitivity analysis of the system was performed with respect to parameters specific to the trailer. The parameters include the mass $m_s$, center of gravity height $h_s$, moment of inertia $\bm{I}$, and cornering stiffness $C_\alpha$.
Additional parameters were investigated but found to have less impact on the system and were therefore omitted.
The resulting sensitivities are presented for a scenario in which the vehicle should maintain a constant $v_x$ of $50$~km/h and linearly increase $\delta_f$ from $0^\circ$ to $8^\circ$. 
The simulations were run for $5$~s; the magnitude reflecting a typical choice of horizon for a predictive controller \cite{fors2022resilient}.
The sensitivity trajectories are multiplied by their respective parameter, illustrating a perturbation of $100$\% of the nominal value.

\begin{figure}[t]
	\includegraphics{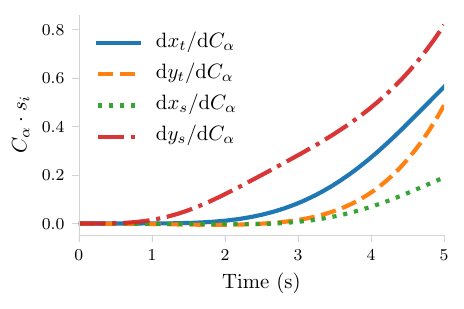}
	\caption{The vehicle planar sensitivity approximations with respect to the trailer cornering stiffness.}
	\label{fig:sa_ca}
\end{figure}

The cornering stiffness is one of the most important model parameters for accurate prediction of lateral position.
Its difficulty to measure or estimate online has further made it a target for investigation in works on robust control \cite{tai1999robust, sadeghi2019robust}.
In \cref{fig:sa_ca}, the sensitivities of the vehicle planar states with respect to the trailer cornering stiffness are presented.
Unsurprisingly, as the steering angle increases, the sensitivities increase as well. 
What is most interesting from these results, however, is that the tractor longitudinal position $x_t$ is seemingly more sensitive than its respective lateral position $y_t$.
This behavior showed strong correlation with the vehicle speed; for lower speeds and greater steering angles, the opposite behavior was recorded. 

\begin{figure}[t]
	\includegraphics{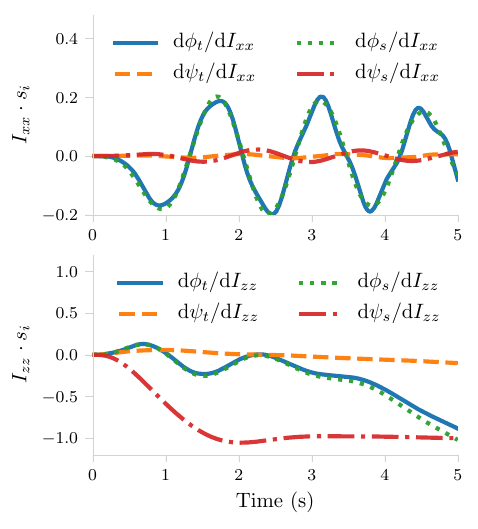}
	\caption{Roll and yaw angle sensitivity approximations ($^\circ$) with respect to the trailer moment of inertia diagonal elements $I_{xx,s}$ and $I_{zz,s}$.}
	\label{fig:sa_i}
\end{figure}

In \cref{fig:sa_i}, the sensitivity trajectory of the vehicle's roll and yaw angles with respect to the trailer moment of inertia is presented.
Since the moment of inertia tensors are expressed in the respective vehicle sprung frame, there is an expected cross-correlation when the vehicle starts rolling or pitching. 
This phenomenon is clearly visible when studying the sensitivity toward $I_{zz,s}$.
As the front wheel steering angle increases, the vehicle will develop a larger roll angle, affecting the sensitivity trajectory. 
The periodicity in the roll angle is explained by the dynamics of the suspension system. 
When the vehicle rolls, the suspension system responds by pulling the vehicle back, briefly alleviating the sensitivity. 

\begin{figure}[t]
	\includegraphics{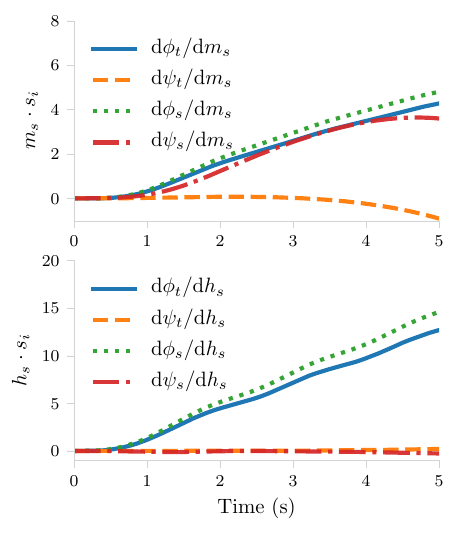}
	\caption{Roll and yaw angle sensitivity approximations ($^\circ$) with respect to the trailer mass $m_{s}$ and COG height $h_{s}$.}
	\label{fig:sa_mh}
\end{figure}

In \cref{fig:sa_mh}, the sensitivity trajectory of the vehicle's roll and yaw angles with respect to the trailer mass and COG height is presented.
It is clear from the results that both of these parameters have significant impact on the sensitivity trajectories. 
The trajectories with respect to $h_s$ precisely illustrate the vulnerability toward minor changes in the trailer dimensions and their impact on the roll angle.

\subsection{Closed-Loop Prediction}
In a real-world setting, it is likely that several model parameter values are uncertain and so it is important to study how a combination of perturbations affect the predictions.
To evaluate how these affect the predictions, we construct two real-world inspired self-driving scenarios:
\begin{itemize}
	\item \emph{Increasing curve} (IC). The vehicle enters a long curve at $45$~km/h, requiring an increasing steering angle. 
	\item \emph{Overtake} (OT). The vehicle performs an overtake starting at $65$~km/h and accelerating to $85$~km/h. 
\end{itemize}
The simulations were first run in closed-loop using nominal parameter values.
The control inputs were stored, and used to evaluate $100$ different disturbance model combinations.
The disturbance models were re-initialized with nominal state values every second and simulated forward over a $5$~s horizon using the stored input commands, illustrating how uncertainties in state predictions would be affected when using a predictive controller.
In order to create the disturbance models, a set of parameters was selected for investigation (see \cref{tab:cl-loop}).
For each investigated parameter, a uniform distribution was then constructed 
\begin{equation}
	X \sim \mathcal{U}(\mu - \epsilon\mu, \mu + \epsilon\mu),
\end{equation}
where $\mu$ is the nominal parameter value and $\epsilon$ a percentage by which the parameter is perturbed, $\epsilon\in \{5, 10, 15\}\%$.

\begin{table}[t]
	\caption{Closed-loop perturbed parameters.}
	\label{tab:cl-loop}
	\centering
	\begin{tabular}{c l l}
		\toprule
		Parameter & Description & Nominal value\\
		\midrule
		$m_s$ & Mass & 32 000 kg\\ 
		$h_s$ & COG height & 1.95 m\\
		$I_{xx,s}$ & Moment of inertia & 4.4e4 kg m$^2$\\
		$I_{zz,s}$ & Moment of inertia & 1.5e5 kg m$^2$\\
		$C_{\alpha_s}$ & Cornering stiffness & 3.3e5 N/rad \\
		$\sigma_s$ & Relaxation length & 0.3 m \\
		\bottomrule
	\end{tabular}
\end{table}

\begin{figure}[t]
	\includegraphics{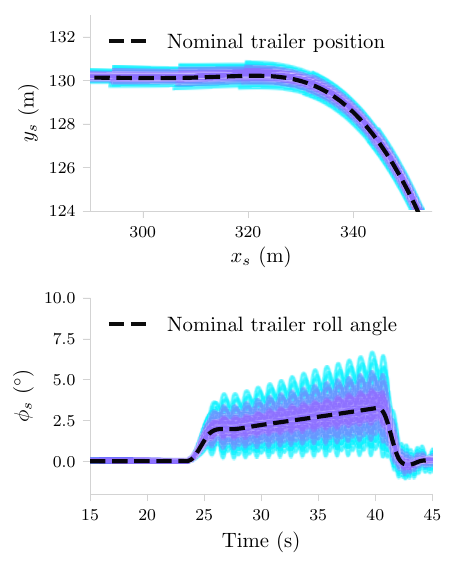}
	\caption{State variable trajectories for the IC scenario using different disturbance models.}
	\label{fig:cl_ic}
\end{figure}

In \cref{fig:cl_ic}, the closed-loop predictions of the trailer position and roll angle are presented for the IC scenario.
The plots illustrate how the predictions deviate from the nominal prediction at increasing degrees of uncertainties. %
Furthermore, the figure showing the trailer position has been zoomed in around the final part of the curve where the curvature is at its largest value.
As can be seen from the results, the uncertainty in position is not necessarily the most critical problem, although the worst-case illustrates a displacement error of $\approx\pm0.5$~m.
The roll angle of the trailer, however, reaches hazardous magnitudes; well above the rollover threshold ($ \approx4^\circ$)---and for several unfortunate combinations also significant oscillations.
Even at modest speeds, the vehicle can enter critical operating regions for minor perturbations in the parameter values.

\begin{figure}[t]
	\includegraphics{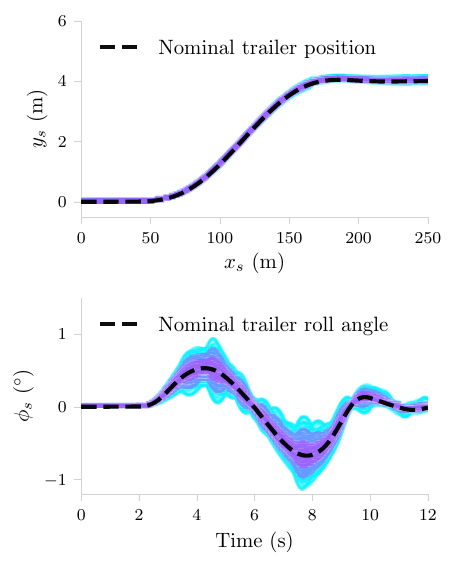}
	\caption{State variable trajectories for the OT scenario using different disturbance models.}
	\label{fig:cl_ot}
\end{figure}

In \cref{fig:cl_ot}, the results for the OT scenario are presented. 
Despite reaching seemingly high speeds, the position of the trailer is not as affected by perturbations in this case compared to the IC scenario.
The decreased spread in position predictions indicate that the steering angle magnitude, determined by the path curvature, has greater impact on position uncertainty than speed.

\subsubsection{Rollover Mitigation Using Disturbance Models}
\begin{figure}[t]
	\includegraphics{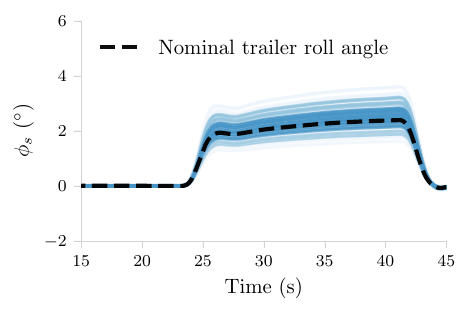}
	\caption{Trailer roll angle for the IC scenario when using the disturbance models' predictions for speed adaptation.}
	\label{fig:cl_clic}
\end{figure}

The results have illustrated that the trailer roll angle is highly susceptible to perturbations in the parameter values.
When model parameter values are uncertain, the predictions of the disturbance models could be used to signal if the vehicle would enter possibly critical situations given the currently planned trajectory.
This concept was evaluated for the IC scenario, with the function to request deceleration if the trailer roll angle deviates too much from a zero-reference, yielding the results in \cref{fig:cl_clic}.
Interestingly, all of the oscillations seen in \cref{fig:cl_ic} can be eliminated, and the angle kept within moderate bounds. 

		\section{Conclusions}
We have implemented and evaluated an 8-DOF model of a tractor-semitrailer vehicle.
The model was used to conduct a sensitivity analysis of the system, providing insight into the relative importance of nominal parameter value accuracy.
We constructed a distribution of model parameters and used sampling-based strategies to compute state variable trajectory uncertainty bounds.
We propose our methods to be used within a robust predictive controller, or as part of a driver--assistance system for rollover or lane departure warning.

\section*{Acknowledgment}
This research was partially supported by the Strategic Reseach Area at
Linköping-Lund in Information Technology (ELLIIT), and partially
supported by the Wallenberg AI, Autonomous Systems and Software
Program (WASP) funded by the Knut and Alice Wallenberg Foundation.

		\bibliographystyle{IEEEtran}
		\bibliography{IEEEabrv,references.bib}{}

\begin{thebibliography}{10}
\providecommand{\url}[1]{#1}
\csname url@samestyle\endcsname
\providecommand{\newblock}{\relax}
\providecommand{\bibinfo}[2]{#2}
\providecommand{\BIBentrySTDinterwordspacing}{\spaceskip=0pt\relax}
\providecommand{\BIBentryALTinterwordstretchfactor}{4}
\providecommand{\BIBentryALTinterwordspacing}{\spaceskip=\fontdimen2\font plus
\BIBentryALTinterwordstretchfactor\fontdimen3\font minus
  \fontdimen4\font\relax}
\providecommand{\BIBforeignlanguage}[2]{{%
\expandafter\ifx\csname l@#1\endcsname\relax
\typeout{** WARNING: IEEEtran.bst: No hyphenation pattern has been}%
\typeout{** loaded for the language `#1'. Using the pattern for}%
\typeout{** the default language instead.}%
\else
\language=\csname l@#1\endcsname
\fi
#2}}
\providecommand{\BIBdecl}{\relax}
\BIBdecl

\bibitem{fors2022resilient}
V.~Fors, B.~Olofsson, and E.~Frisk, ``Resilient branching {MPC} for
  multi-vehicle traffic scenarios using adversarial disturbance sequences,''
  \emph{IEEE Transactions on Intelligent Vehicles}, vol.~7, no.~4, pp.
  838--848, 2022.

\bibitem{westny2023graph}
T.~Westny, J.~Oskarsson, B.~Olofsson, and E.~Frisk, ``{MTP-GO}: Graph-based
  probabilistic multi-agent trajectory prediction with neural {ODEs},''
  \emph{IEEE Transactions on Intelligent Vehicles}, vol.~8, no.~9, pp.
  4223--4236, 2023.

\bibitem{zhou2023interaction}
J.~Zhou, B.~Olofsson, and E.~Frisk, ``Interaction-aware motion planning for
  autonomous vehicles with multi-modal obstacle uncertainty predictions,''
  \emph{IEEE Transactions on Intelligent Vehicles}, 2023.

\bibitem{sampson2000active}
D.~J.~M. Sampson, ``Active roll control of articulated heavy vehicles,'' Ph.D.
  dissertation, Cambridge University Engineering Department, 2000.

\bibitem{tai1999robust}
M.~Tai and M.~Tomizuka, ``Robust lateral control of heavy duty vehicles for
  automated highway systems,'' \emph{IFAC Proceedings Volumes}, vol.~32, no.~2,
  pp. 8309--8314, 1999.

\bibitem{sadeghi2019robust}
M.~Sadeghi~Kati, H.~K{\"o}ro{\u{g}}lu, and J.~Fredriksson, ``Robust lateral
  control of long-combination vehicles under moments of inertia and tyre
  cornering stiffness uncertainties,'' \emph{Vehicle System Dynamics}, vol.~57,
  no.~12, pp. 1847--1873, 2019.

\bibitem{kong2015kinematic}
J.~Kong, M.~Pfeiffer, G.~Schildbach, and F.~Borrelli, ``Kinematic and dynamic
  vehicle models for autonomous driving control design,'' in \emph{2015 IEEE
  Intelligent Vehicles Symposium (IV)}, 2015, pp. 1094--1099.

\bibitem{islam2019well}
M.~M. Islam, N.~Fr{\"o}jd, S.~Kharrazi, and B.~Jacobson, ``How well a
  single-track linear model captures the lateral dynamics of long combination
  vehicles,'' \emph{Vehicle System Dynamics}, vol.~57, no.~12, pp. 1874--1896,
  2019.

\bibitem{gafvert20049}
M.~G{\"a}fvert and O.~Lindg{\"a}rde, ``A 9-dof tractor-semitrailer dynamic
  handling model for advanced chassis control studies,'' \emph{Vehicle System
  Dynamics}, vol.~41, no.~1, pp. 51--82, 2004.

\bibitem{chen2000lateral}
C.~Chen and M.~Tomizuka, ``Lateral control of commercial heavy vehicles,''
  \emph{Vehicle System Dynamics}, vol.~33, no.~6, pp. 391--420, 2000.

\bibitem{wong2008theory}
J.~Y. Wong, \emph{Theory of ground vehicles}.\hskip 1em plus 0.5em minus
  0.4em\relax John Wiley \& Sons, 2008.

\bibitem{pacejka2012tire}
H.~Pacejka, \emph{Tire and vehicle dynamics}, 3rd~ed.\hskip 1em plus 0.5em
  minus 0.4em\relax Elsevier, 2012.

\bibitem{ljung1999system}
L.~Ljung, \emph{System identification: Theory for the User}, 2nd~ed.\hskip 1em
  plus 0.5em minus 0.4em\relax Prentice Hall, 1999.

\bibitem{maly1996numerical}
T.~Maly and L.~R. Petzold, ``Numerical methods and software for sensitivity
  analysis of differential-algebraic systems,'' \emph{Applied Numerical
  Mathematics}, vol.~20, no.~12, pp. 57--79, 1996.

\bibitem{shampine1997matlab}
L.~F. Shampine and M.~W. Reichelt, ``The matlab ode suite,'' \emph{SIAM Journal
  on Scientific Computing}, vol.~18, no.~1, pp. 1--22, 1997.

\bibitem{paden2016survey}
B.~Paden, M.~{\v{C}}{\'a}p, S.~Z. Yong, D.~Yershov, and E.~Frazzoli, ``A survey
  of motion planning and control techniques for self-driving urban vehicles,''
  \emph{IEEE Transactions on Intelligent Vehicles}, vol.~1, no.~1, pp. 33--55,
  2016.

\end{thebibliography}

	\end{document}